\documentclass[twocolumn,showpacs,floatfix]{revtex4}
\usepackage{graphics,epsf,bm}

\def\ninj#1#2#3#4#5#6#7#8#9{\left\{ \begin{array}{ccc}
                    #1 & #2 & #3 \\ #4 & #5 & #6 \\
                    #7 & #8 & #9  \end{array} \right\}}

\def\bK{\mbox{\boldmath $K$}}

\def\bQ{\mbox{\boldmath $Q$}}
\def\bR{\mbox{\boldmath $R$}}

\def\bk{\mbox{\boldmath $k$}}

\def\bp{\mbox{\boldmath $p$}}
\def\bq{\mbox{\boldmath $q$}}
\def\br{\mbox{\boldmath $r$}}
\def\bs{\mbox{\boldmath $s$}}

\begin{document}
\title{Localized $N, \Lambda, \Sigma$, and $\Xi$ Single-Particle Potentials
in Finite Nuclei Calculated with $SU_6$ Quark-Model Baryon-Baryon Interactions}
\author{M. Kohno$^1$ and Y. Fujiwara$^2$}
\affiliation{$^1$Physics Division, Kyushu Dental College, Kitakyushu
803-8580, Japan\\
$^2$Department of Physics, Kyoto University, Kyoto 606-8502, Japan\\
}

\begin{abstract}
Localized single-particle potentials for all octet baryons, $N$, $\Lambda$,
$\Sigma$, and $\Xi$, in finite nuclei, $^{12}$C, $^{16}$O, $^{28}$Si,
$^{40}$Ca, $^{56}$Fe, and $^{90}$Zr, are calculated using the quark-model
baryon-baryon interactions. $G$-matrices evaluated in symmetric nuclear
matter in the lowest order Brueckner theory are applied to finite nuclei
in local density approximation. Non-local potentials are localized by
a zero-momentum Wigner transformation. Empirical single-particle properties
of the nucleon and the $\Lambda$ hyperon in nuclear medium have been known
to be explained semi-quantitatively in the LOBT framework.
Attention is focused in the present consideration on predictions
for the $\Sigma$ and $\Xi$ hyperons. The unified description for
the octet baryon-baryon interactions by the SU$_6$ quark-model enables us to
obtain less ambiguous extrapolation to the $S=-1$ and $S=-2$ sectors
based on the knowledge in the $NN$ sector than other potential models.
The $\Sigma$ mean field is shown to be weakly attractive at the surface,
but turns to be repulsive inside, which is consistent with the experimental
evidence. The $\Xi$ hyperon s.p. potential is also attractive at the nuclear
surface region, and inside fluctuates around zero. Hence $\Xi$ hypernuclear
bound states are unlikely. We also evaluate energy shifts of the $\Sigma^-$
and $\Xi^-$ atomic levels in $^{28}$Si and $^{56}$Fe, using the calculated
s.p. potentials.
\end{abstract}

\maketitle

\section{Introduction}
One of the salient features of atomic nuclei is the success of the description
of their properties by a single-particle (s.p.) picture. Although the
nucleon-nucleon interaction is known to be strongly repulsive at the short
range part, which was once conveniently described by a hard-core, the nucleon
single-particle potential is well represented by a well-behaved local potential
of the Woods-Saxon form. The theoretical base of understanding this circumstance
in view of the singular two-body interaction was provided by the Brueckner
theory in 1950's \cite{LOBT1,LOBT2,DAY}. The progress of the density-dependent
Hartree-Fock (DDHF) description of nuclear bulk properties followed
in 1970's \cite{NEG,CS}, introducing some phenomenological adjustment
for $G$-matrices in the Brueckner theory. 

The mean field picture seems to hold also for hyperons in nuclei. For the
$\Lambda$ hyperon, the potential properties have been known from light to
heavy nuclei from $\Lambda$ formation and spectroscopy experiments \cite{HT}.
Experimental studies of the $\Sigma$ and $\Xi$ hyperons in
nuclear medium properties are now in progress. Because direct hyperon-nucleon
scattering experiments are not readily available, the properties of the hyperon s.p.
potentials are a valuable source of hyperon-nucleon interactions. This case, we have
to resort to an effective interaction theory to relate s.p. properties of the hyperon
embedded in nuclei with the character of hyperon-nucleon interactions.

In this paper we develop a method to obtain local potentials for octet baryons
in finite nuclei with using full non-local $G$-matrix elements in nuclear matter,
starting from the baryon-baryon bare interactions.
The calculation of single-particle properties in nuclear matter can provide the
basic information about the baryons in nuclear medium derived from the bare
interaction. Nevertheless, it is instructive to explicitly calculate the s.p. potential
in finite nuclei starting from two-body baryon-baryon interactions and
compare them with the empirical ones. It is not even obvious whether the shape
represented by the Woods-Saxon form which has been established both for the
nucleon and the $\Lambda$ hyperon mean fields is suitable for the $\Sigma$
and the $\Xi$ hyperons. The straightforward folding of
the two-body effective interaction in momentum space provides a non-local potential
in a nucleus. The non-locality also comes from the exchange character of the basic
interaction. In order to make a comparison with empirical data, it is meaningful to
define a local potential by some localization procedure.
We employ in this paper a zero-momentum Wigner transformation
method based on the WKB localization approximation \cite{HH}.

It is necessary for a predictive discussion about hyperon s.p. properties in nuclei
to use octet baryon-baryon bare interactions as reliable as possible.
With little experimental information except for the $\Lambda N$ interaction,
the construction of the interactions in the strangeness $S=-1$ and $S=-2$
sectors is not simple, although some constraints are imposed by the flavor
symmetry. The typical potential model has been developed in a one-boson-exchange
potential (OBEP) picture by
the Nijmegen group. The early parametrization with the hard-core
in the 1970s \cite{HCD,HCF} has been successively revised by adjusting
parameters in the soft core version \cite{NSC,NSC99} and introducing new
terms \cite{ESC06}. There are now a number of sets of parameters, reflecting
ambiguities due to the lack of experimental data. Although the description for
the $\Lambda N$ interaction seems to be under control, there are
various uncertainties in the $\Sigma N$ sector. For the $\Xi$ hyperon, namely
the strangeness $S=-2$ sector, the situation is no better.

Using the spin-flavor SU$_6$ quark-model, the Kyoto-Niigata
group \cite{FU95,FU96a,FU96b,FSS2,QMBB} have developed
a unified description for the interactions between full octet baryons. In this model,
the interaction is constructed as the Born kernel in the framework of the resonating
group method for the three-quark clusters, the short range part of which is composed
of the effective one-gluon exchange mechanism. The basic SU$_6$ symmetry
provides a specific framework to the interactions between octet baryons and the
Pauli principle respected on the quark level in addition  brings about
a characteristic structure to them. Incorporating effective meson exchange
potentials between quarks, namely the scalar, pseudo-scalar, and vector mesons
exchanges, the model is able to account the $NN$ scattering data as accurately
as other modern $NN$ potential models.

Parameters of the SU$_6$ quark-model by the Kyoto-Niigata group are mostly fixed
in the $NN$ and $\Lambda N$ sectors, and the uncertainties in the extension to
the $\Sigma N$ and $\Xi N$ channels are limited. In fact, the definite predictions
such as the smallness of the $\Lambda N$ spin-orbit interaction and the overall
repulsive nature of the $\Sigma$-nucleus s.p. potential in nuclei have been
supported by the experiments afterwords.
Therefore it is interesting to examine the whole prediction
of this potential for s.p. properties of all the octet baryons in nuclear medium.
In particular, concrete predictions are presented for the $\Xi$ hyperon.
We use the most recent quark-model potential fss2 \cite{FSS2,QMBB} in this paper.
The parameter set includes no adjustable parameter
for the tuning afterward. The original
interaction as a Born kernel has an inherent energy-dependence. Recently
the method to eliminate the energy-dependence was developed \cite{SUZ08}.
We actually use this renormalized version of the fss2 potential.

We present, in Sec. II, basic expressions of the method to evaluate localized
$N$, $\Lambda$, $\Sigma$, and $\Xi$ s.p. potentials in a finite nucleus. 
We first discuss numerical results of these s.p. potentials in nuclear matter
to represent basic characters of the $G$-matrices of the
quark-model baryon-baryon interactions.
Calculated results in finite nuclei, $^{12}$C, $^{16}$O, $^{28}$Si, $^{40}$Ca,
$^{56}$Fe, and $^{90}$Zr, are shown in Sec. IV. The energy shift and the width
of the $\Sigma^-$ and $\Xi^-$ atomic levels in $^{28}$Si and $^{56}$Fe are
studied in Sec. V on the basis of the s.p. potential obtained in Sec. IV.
Summary is given in Sec. VI. 

\section{Localized single-particle potential in a finite nucleus}
We calculate a baryon single-particle potential in a finite nucleus which is
defined by folding $NN$ or $YN$ $G$-matrix elements in nuclear matter with
respect to nucleon occupied states through the local density approximation.
It has been a traditional method for the microscopic study of bulk properties
of nuclei to construct density-dependent two-body local interaction based on
the $G$-matrices in nuclear matter and apply the effective interaction to mean
field calculations for finite nuclei. Avoiding this procedure, we directly
fold $G$-matrix elements to obtain non-local s.p. potentials and localize them.
In the DDHF calculations, some phenomenological adjustments are introduced to
reproduce the properties of the well-known nuclei. The purpose of the present
paper is not to accomplish the reproduction of empirical values, but examine
overall implications of the unified description of the baryon-baryon bare
interaction by the quark model potential fss2 \cite{QMBB} for hyperon s.p.
potentials in finite nuclei. In this section, we derive a basic expression
for the s.p. potential, by introducing some approximations
and a localized method by the zero-momentum Wigner transform.

\subsection{direct term}
First we consider the following direct term contribution. The wave function
$\phi_{\ell_h j_h m_{j_h}}$ denotes the nucleon s.p. wave function of the nucleus
with the orbital angular momentum $\ell_h$ and the total angular momentum $j_h$,
and the dummy wave function of the baryon for which the potential is calculated
is denoted by $\phi_{\ell j m_j}$. The average over the $z$-component of
the total angular momentum, $\frac{1}{\hat{j}}\sum_{m_j}$
with $\hat{j}\equiv 2j+1$, means that we assume the spherical symmetry from
the beginning. We do not write the isospin indices for simplicity in the
following derivation, but recover them in the final expression.
\begin{widetext}
\begin{eqnarray}
I_D &\equiv & \frac{1}{\hat{j}} \sum_{m_j} \sum_{h m_{j_h}}
 \int\!\!\int\!\!\int\!\!\int d\br_1d\br_2d\br_1'd\br_2'\;
 \phi_{\ell j m_j}^*(\br_1') \phi_{\ell_h j_h m_{j_h}}^*(\br_2')
 G(\br_1',\br_2',\br_1,\br_2)
 \phi_{\ell j m_j}(\br_1) \phi_{\ell_h j_h m_{j_h}}(\br_2) \nonumber \\
 &=& \sum_{h} \sum_{JMLL'S} \hat{j_h}\hat{S} \sqrt{\hat{L}\hat{L'}}
 \ninj{\ell}{\ell_h}{L}{1/2}{1/2}{S}{j}{j_h}{J}
 \ninj{\ell}{\ell_h}{L'}{1/2}{1/2}{S}{j}{j_h}{J}
 \int\!\!\int\!\!\int\!\!\int d\br_1d\br_2d\br_1'd\br_2' \nonumber \\
{} & & \times [[\phi_\ell^*(\br_1')\times
 \phi_{\ell_h}^*(\br_2')]^{L'}\times \chi^S ]^J_M G(\br_1',\br_2',\br_1,\br_2)
[[\phi_\ell (\br_1)\times\phi_{\ell_h}(\br_2)]^{L}\times \chi^{S} ]^{J}_{M}.
\end{eqnarray}
The effective baryon-baryon interaction $G(\br_1',\br_2',\br_1,\br_2)$ in a
coordinate space is supposed to be related to the $G$-matrix in momentum
space $G(\bk',\bk;K,\omega)$ by
\begin{equation}
G(\br_1',\br_2',\br_1,\br_2) = \frac{(2\pi)^3}{(2\pi)^{12}}
\int\!\!\int\!\!\int\!\!\int d\bk_1d\bk_2d\bk_1'd\bk_2' \delta(\bK -\bK')
 e^{i(\bk_1'\cdot \br_1'+\bk_2'\cdot \br_2'-\bk_1\cdot \br_1-\bk_2\cdot \br_2)}
 G(\bk',\bk;K,\omega),
\end{equation}
Here each momentum has the following relation: $\bk_1=\frac{m_1}{m_1+m_2}\bK+\bk$,
$\bk_2=\frac{m_2}{m_1+m_2}\bK-\bk$,  $\bk_1'=\frac{m_1}{m_1+m_2}\bK'+\bk'$,
$\bk_2'=\frac{m_2}{m_1+m_2}\bK'-\bk'$. The mass of the baryon for which the
s.p. potential is evaluated is denoted by $m_1$ and the mass of the nucleon in
the target nucleus  by $m_2$. The $G$-matrix is evaluated in symmetric nuclear
matter by solving the baryon-channel coupling Bethe-Goldstone equations
\begin{equation}
 G_{\alpha,\alpha'}(\bk',\bk;K,\omega) = V_{\alpha,\alpha'}(\bk',\bk;\bK)
 + \frac{1}{(2\pi)^3} \sum_\beta \int d\bq V_{\alpha,\beta}(\bk',\bq;\bK)
 \frac{Q_\beta(q,K)}{\omega -E_b(k_1)-E_N(k_2)}G_{\beta,\alpha'}(\bq,\bk;K,\omega),
\end{equation}
\end{widetext}
with the suffix specifying the pair of a baryon $b$ and a nucleon $N$ by $\beta$.
The Pauli exclusion operator $Q_\beta$ is treated in the standard angle-average
approximation. The explicit expression may be found in ref. \cite{KF00}. $E_a(k)$
is a s.p. energy of the baryon $a$ in nuclear matter. We employ the continuous
choice for the energy denominator in Eq. (3).
That is, $E_a(k)=m_a + \frac{\hbar^2}{2m_a}k^2+U_a(k)$ is defined
self-consistently by the following definition of the s.p. potential $U_a$.
\begin{equation}
 U_a(k)= \int d\bk' G_{\alpha,\alpha}(\bq, \bq; \bk+\bk', \omega),
\end{equation}
where $\bq = \frac{1}{2}(\bk-\bk')$ and $\omega = E_{N,Y}(k)+E_N(k')$.
The prescription for the starting energy $\omega$ in the local density approximation
is explained in the next section.

The straightforward calculation of Eq. (1) needs much computational effort and is
not fruitful to obtain a physical insight for baryon properties in nuclei by
starting from the bare baryon-baryon interactions. We introduce two simplifying
approximations. One is the spin-average in taking the sum of the matrix elements,
which means that we take the average over the spin orientation: 
\begin{eqnarray}
 & & \frac{1}{\hat{S}} \sum_{M_S} \langle SM_S|G (\bk',\bk;K,\omega)|SM_S\rangle
 \nonumber \\
 &=& \sum_{q J_q} \frac{\hat{J_q}}{4\pi\hat{J}_q\hat{S}} G_{qq}^{J_qS} (k',k;K,\omega)
 P_q (\cos \widehat{\bk'\bk}),
\end{eqnarray}
where the $G$-matrix is decomposed to partial waves and $P_q$ stands for the
Legendre polynomial with $q$ specifying the orbital angular momentum.
The other simplification is the following replacement.
\begin{eqnarray}
& & \int\!\!\int d\br_1 d\br_1' \phi_{\ell j m_j}^*(\br_1')\phi_{\ell j m_j}(\br_1)
 \nonumber \\ 
 &=& \int\!\!\int d\bR_1d\bs_1 
\phi_{\ell j m_j}^*(\bR_1+\frac{1}{2}\bs_1)\phi_{\ell j m_j}(\bR_1-\frac{1}{2}\bs_1)
 \nonumber \\
 &\Rightarrow& \int d\bR_1\phi_{\ell j m_j}^*(\bR_1)\phi_{\ell j m_j}(\bR_1) \int d\bs_1.
\end{eqnarray}
This corresponds to the zero-momentum Wigner transformation of the non-local potential.
That is, we set $p=0$ for the Wigner transformation $U^{W}(\bR,\bp)$ of the non-local
potential $U(\br_1,\br_2)$.
\begin{equation}
 U^{W}(\bR,\bp) =\int d\bs e^{i\bp\cdot \bs} U(\bR+\frac{1}{2}\bs,\bR-\frac{1}{2}\bs).
\end{equation}
Results shown in Sect. IV for nucleons and lambdas, for which we know what s.p.
potentials potentials are expected in $G$-matrix calculations with bare $NN$
and $\Lambda N$ interactions in various studies in literature, implies that
the zero-momentum approximation works well. More direct confirmation of the
reliability of this approximation will be presented elsewhere \cite{ZMA}.

To evaluate Eq. (1) with the above simplification it is convenient to use the
Fourier transform of the s.p. wave function $\phi_{\ell jm_j}$,
\begin{eqnarray}
\tilde{\phi}_{\ell jm_j} (\bk)&=&\frac{1}{(2\pi)^3} \int d\br e^{-i\bk\cdot\br} \phi(\br)
 \nonumber \\
 &=& \frac{1}{(2\pi)^{3/2}} i^{2n-\ell} [Y_\ell (\hat{\bk})\times \chi_{1/2} ]^j_{m_j}
 \frac{1}{k} \tilde{\phi}_{\ell j}(k),
\end{eqnarray}
where $n$ is a nodal quantum number and the Fourier transformation of the radial wave
function is defined as
\begin{equation}
 \frac{1}{k}  \tilde{\phi}_{\ell j}(k)= (-i)^{2n} \sqrt{\frac{2}{\pi}}
 \int dr\: r j_\ell (kr) \phi_{\ell j}(r).
\end{equation}
After carrying out some integrations and taking angular momentum recouplings,
we obtain the final expression as follows.
\begin{eqnarray}
I_D &=& \frac{1}{4(4\pi)^2} \frac{1}{(2\pi)^3} \left( 1+\frac{m_2}{m_1} \right)^3
 \sum_{h} \hat{j_h}\int dR_1 |\phi_{\ell}(R_1)|^2  \nonumber \\
 & \times &\! \int\!\!\int d\bk d\bk'  j_0(|\bk'-\bk|R_1)
 \frac{1}{|\bQ_1'|} \tilde{\phi}_{\ell_h}^* (|\bQ_1'|)  \nonumber \\
 &\times& \frac{1}{|\bQ_1|} \tilde{\phi}_{\ell_h} (|\bQ_1|)
 P_{\ell_h} (\cos \widehat{\bQ'\bQ}) \nonumber \\
& \times& \sum_{q J_qS} \hat{J_q} G_{qq}^{J_qS}(k,k') P_q (\cos \widehat{\bk\bk'}),
\end{eqnarray}
where $\bQ_1$ and $\bQ_1'$ are defined by $\bk$ and $\bk'$ as 
\begin{eqnarray}
 \bQ_1 &\equiv& -\left(1+\frac{m_2}{2m_1}\right)\bk - \frac{m_2}{2m_1}\bk', \\
 \bQ_1' &\equiv& -\left(1+\frac{m_2}{2m_1}\right)\bk' - \frac{m_2}{2m_1}\bk.
\end{eqnarray}

\subsection{Exchange term}
We also have to consider the exchange term contribution, which is familiar
for the nucleon through the antisymmetrization of the wave function.
For hyperons, such terms appear in association with the exchange character
of the hyperon-nucleon interaction, which is realized by the strange
meson exchange in the OBEP description. Denoting the space-exchange operator
and the spin-exchange operator by $P_r$ and $P_\sigma$, respectively,
and specifying the even and odd components of the interaction under
the space-exchange, the matrix element of the $Y N$ interaction is written as
\begin{eqnarray}
 \langle YN|V|YN \rangle &=& \langle YN|V_E\frac{1}{2}(1+P_r)
 \!+\!V_O \frac{1}{2}(1-P_r) |YN\rangle \nonumber \\
 &=& \langle YN |\frac{1}{2}(V_E + V_O)|YN\rangle \nonumber \\
 & &-\langle YN| \frac{1}{2}(V_O -V_E)P_\sigma |NY\rangle,
\end{eqnarray} 
where the relation $P_\sigma P_r |NY\rangle = |YN\rangle$ is used. The first
term was treated in the previous subsection as a direct term contribution,
and the second term is considered in this subsection. The effective interactions
in the direct and exchange contributions should be treated as such a combination
of the even and odd parts in each spin and isospin channels, though the isospin
dependence is disregarded in the above expression because the inclusion of it
in the final expression is simple.
\begin{widetext}
\begin{eqnarray}
I_E &\equiv & -\frac{1}{\hat{j}} \sum_{m_j} \sum_{h m_{j_h}}
 \int\!\!\int\!\!\int\!\!\int d\br_1d\br_2d\br_1'd\br_2'\;
 \phi_{\ell_h j_h m_{j_h}}^*(\br_1') \phi_{\ell j m_j}^*(\br_2')
 G(\br_1',\br_2',\br_1,\br_2)
 \phi_{\ell j m_j}(\br_1) \phi_{\ell_h j_h m_{j_h}}(\br_2) \nonumber \\
 &=& - \sum_{h} \sum_{JMLL'S} \hat{j_h}\hat{S} \sqrt{\hat{L}\hat{L'}} (-1)^{j+j_h-J}
 \ninj{\ell}{\ell_h}{L}{1/2}{1/2}{S}{j}{j_h}{J}
 \ninj{\ell}{\ell_h}{L'}{1/2}{1/2}{S}{j}{j_h}{J} 
 \int\!\!\int\!\!\int\!\!\int d\br_1d\br_2d\br_1'd\br_2' \nonumber \\
{} & & \times (-1)^{\ell_h+\ell+1+j_h+j+L'+S+J}[[\phi_{\ell_h}^*(\br_1')
 \times\phi_{\ell}^*(\br_2')]^{L'}\times \chi^S ]^J_M 
 G(\br_1',\br_2',\br_1,\br_2)
[[\phi_\ell (\br_1)\times\phi_{\ell_h}(\br_2)]^{L}\times \chi^{S} ]^{J}_{M}
\end{eqnarray}
\end{widetext}
In this case we define $\bR_1$ and $\bs_1$ as
\begin{equation} \bR_1=\frac{1}{2}(\br_1+\br_2'),\hspace{1em} \bs_1=\br_2'-\br_1.
\end{equation}
Introducing the same simplifying approximations as in the direct term, we obtain
\begin{eqnarray}
 \! &I_E&\! = \frac{-1}{4(4\pi)^2}\frac{1}{(2\pi)^3}
 \left(1+\frac{m_2}{m_1}\right)^3\sum_{h}
 \hat{j_h}  \int\!dR_1\; |\phi_\ell (R_1)|^2 \nonumber \\
 & \times &\!\!\int\!\!\int \! d\bk d\bk'  j_0 (|\bk+\bk'|R_1) \frac{1}{|\bQ_2'|}
  \tilde{\phi}_{\ell_h}^*(|\bQ_2'|) \nonumber \\
 &\times & \frac{1}{|\bQ_2|} \tilde{\phi}_{\ell_h}(|\bQ_2|)
 P_{\ell_h} (\cos \widehat{\bQ_2'\bQ_2}) \nonumber \\
 & \times & \!\!\!\! \sum_{qJ_qS} (-1)^{1+S} \hat{J_q} G_{qq}^{J_qS}(k,k')
 P_{q} (\cos \widehat{\bk\bk'}),
\end{eqnarray}
where $\bQ_2$ and $\bQ_2'$ are defined by $\bk$ and $\bk'$ by
\begin{eqnarray}
  \bQ_2 &=& -\left(1+\frac{m_2}{2m_1}\right) \bk +\frac{m_2}{2m_1}\bk',\\
  \bQ_2'  &=& \left(1+\frac{m_2}{2m_1}\right) \bk' -\frac{m_2}{2m_1}\bk. \\
  \nonumber
\end{eqnarray}
These $\bQ_2$ and $\bQ_2'$ are obtained by changing the sign of $\bk'$
in $\bQ_1$ and $\bQ_1'$ of Eqs. (11) and (12). It is easy to see that the
difference of the expressions of $I_E$ and $I_D$ is only the factor
$(-1)^{S+q}$. Thus, recovering the isospin degrees of freedom, we obtain
\begin{widetext}
\begin{eqnarray}
 I_D+I_E &=& \frac{1}{4(4\pi)^2} \sum_{hS} \int\!dR_1\; |\phi_\ell (R_1)|^2
 \hat{j_h} \left(1+\frac{m_2}{m_1}\right)^3 \frac{1}{(2\pi)^3} \!\int\!\!\int
 d\bk d\bk' j_0 (|\bk+\bk'|R_1) \frac{1}{|\bQ_2'|} \tilde{\phi}_{\ell_h}^*(|\bQ_2'|)
 \frac{1}{|\bQ_2|} \tilde{\phi}_{\ell_h}(|\bQ_2|) \nonumber \\
{} & & \times     P_{\ell_h} (\cos \widehat{\bQ_2'\bQ_2})
  \sum_{qJ_qS} (1+(-1)^{S+q+I_B+1/2-T}) (I_B M_B 1/2 i_h|T M_B+i_h)^2 
 \hat{J_q} G_{qq}^{J_qST}(k,k') P_{q} (\cos \widehat{\bk\bk'}).
\end{eqnarray}
\end{widetext}

\begin{widetext}
\begin{figure}[t]
\begin{minipage}{0.7\textwidth}
 \epsfxsize=\textwidth
 \epsffile{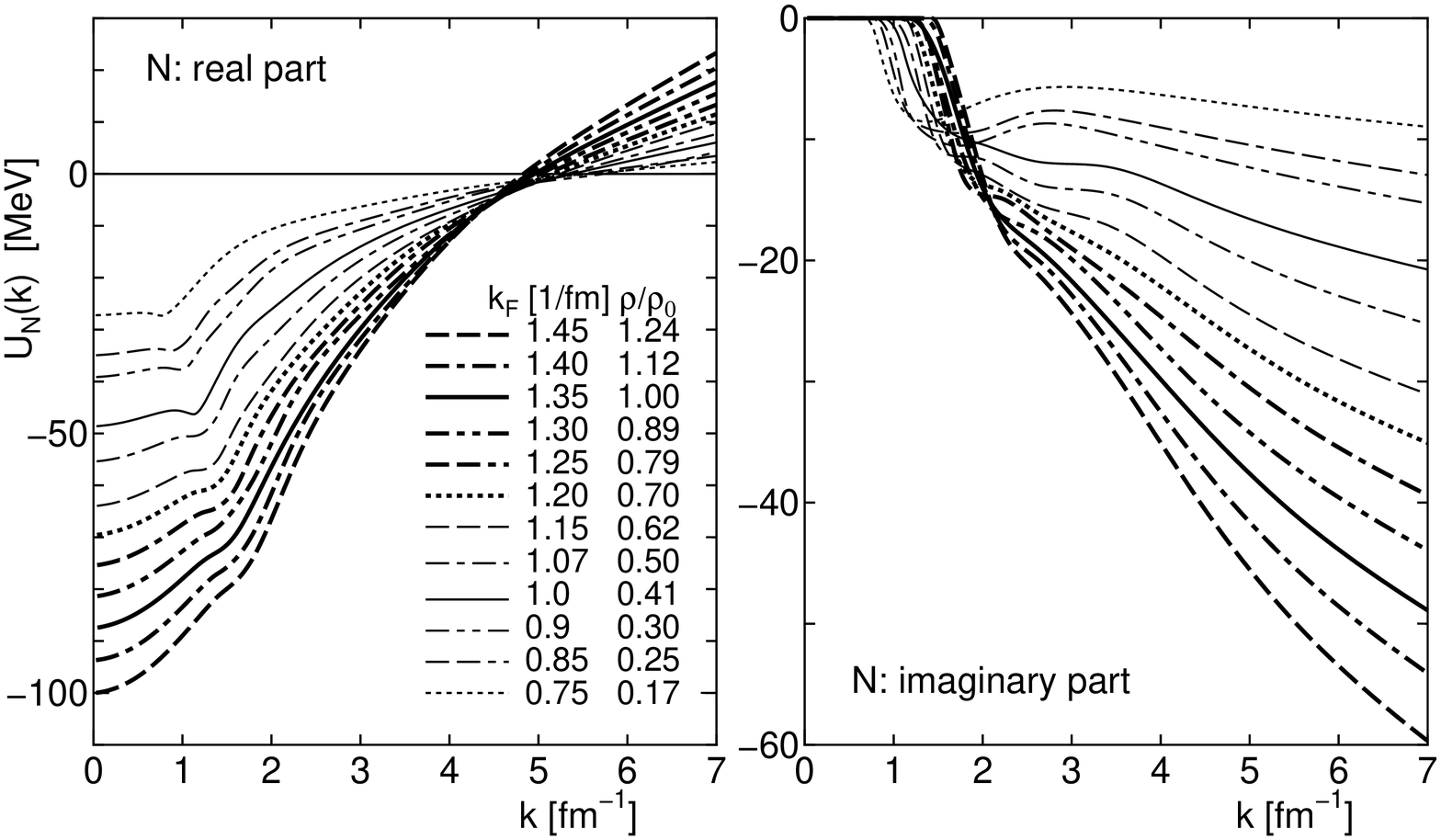}
\\
\medskip
 \epsfxsize=\textwidth
 \epsffile{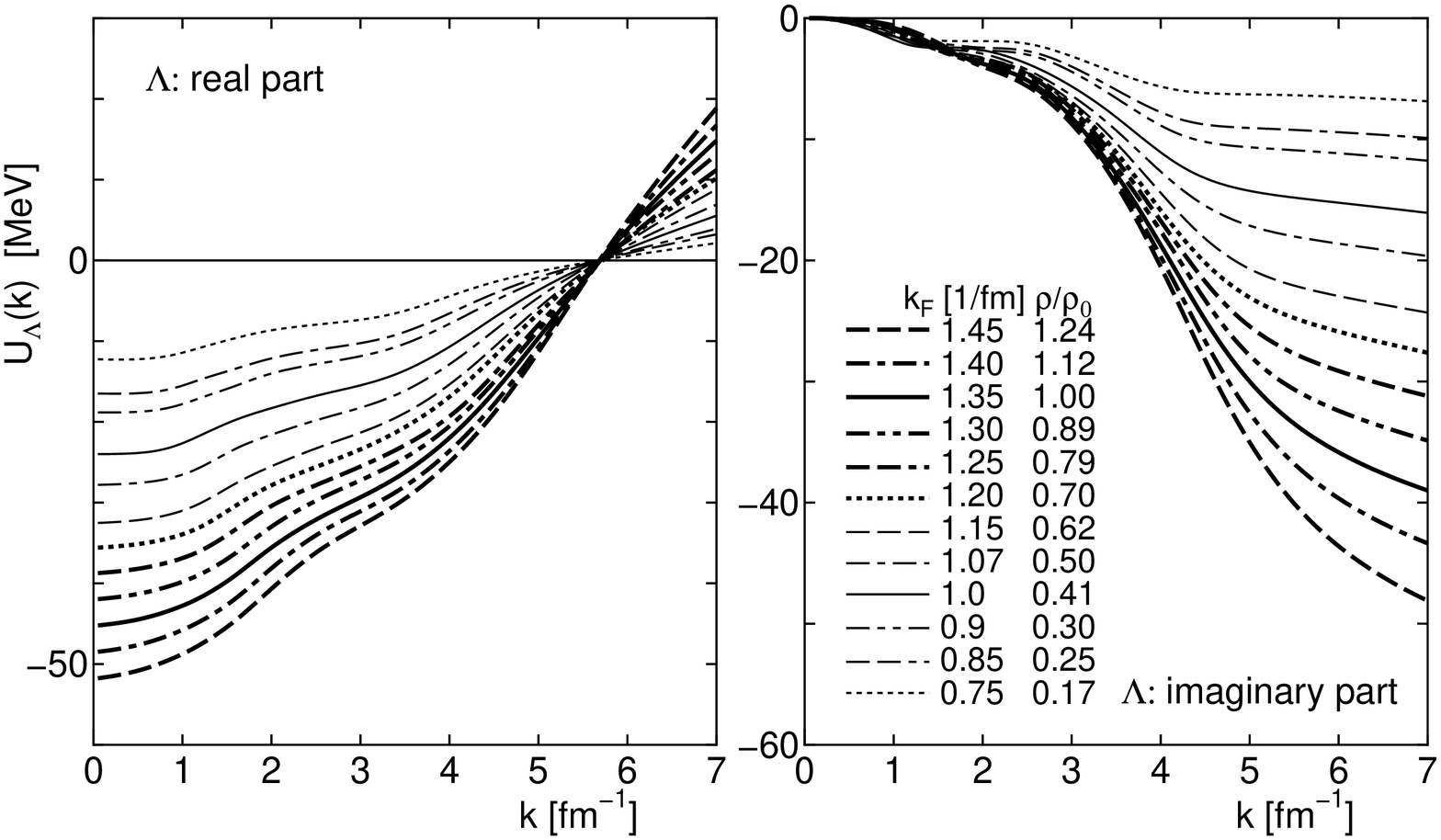}
\end{minipage}
\nolinebreak
\hspace*{5mm}
\begin{minipage}{0.26\textwidth}
 \caption{Real and imaginary parts of single-particle potentials for $N$ and $\Lambda$
in symmetric nuclear matter at various
Fermi momenta, $0.75$ fm$^{-1}$ $\leq k_F \leq$ 1.45 fm$^{-1}$.}
\end{minipage}
\end{figure}
\end{widetext}

In the above expression. $I_B$ is the isospin of the baryon for which the s.p.
potential is considered, and $M_B$ is its $z$-component. The index $i_h$ denotes
the proton or neutron in the target nucleus.

The Eq. (19) defines a s.p. potential $U_B(R)$ to give
\begin{equation}
 I_D+I_E=4\pi \int R^2 dR ~ |\phi_{\ell j}(R)|^2 U_B(R)
\end{equation}
It is noted that the potential $U_B(R)$ does not have $\ell$- and
$j$-dependences due to approximations introduced in the derivation.

\section{Single-particle potentials in Symmetric Nuclear Matter}

Before discussing baryon s.p. potentials in finite nuclei, we show s.p. potentials
in nuclear matter at various Fermi momenta, $0.75 \le k_F \le 1.45$ fm$^{-1}$,
with the quark-model potential fss2 \cite{QMBB}. This potential is defined as a Born
kernel of the RGM description of the interaction between the three-quark clusters.
We use the energy-independent renormalized version of the fss2 potential \cite{SUZ08}.
The details of the $G$-matrix calculation for hyperons in nuclear matter are
reported in ref. \cite{KF00}. It has been known that the LOBT saturation curve
in ordinary nuclear matter does not reproduce the empirical saturation property.
Although the curve obtained by the potential fss2 with the continuous choice for
intermediate spectra almost goes through the empirical saturation point of
$k_F=1.35$ fm$^{-1}$ and $E/A=-16.5$ MeV, the energy minimum appears at
$k_F=1.7$ fm$^{-1}$ and $E/A=-20$ MeV. Nevertheless the LOBT calculation provides
a useful starting point and meaningful information for the baryon s.p. potentials
in nuclear medium in the microscopic studies based on the bare baryon-baryon
interactions. Missing effects  in the LOBT , such as con-
\pagebreak

\begin{widetext}
\begin{figure}[t]
\begin{minipage}{0.7\textwidth}
{ \epsfxsize=\textwidth
 \epsffile{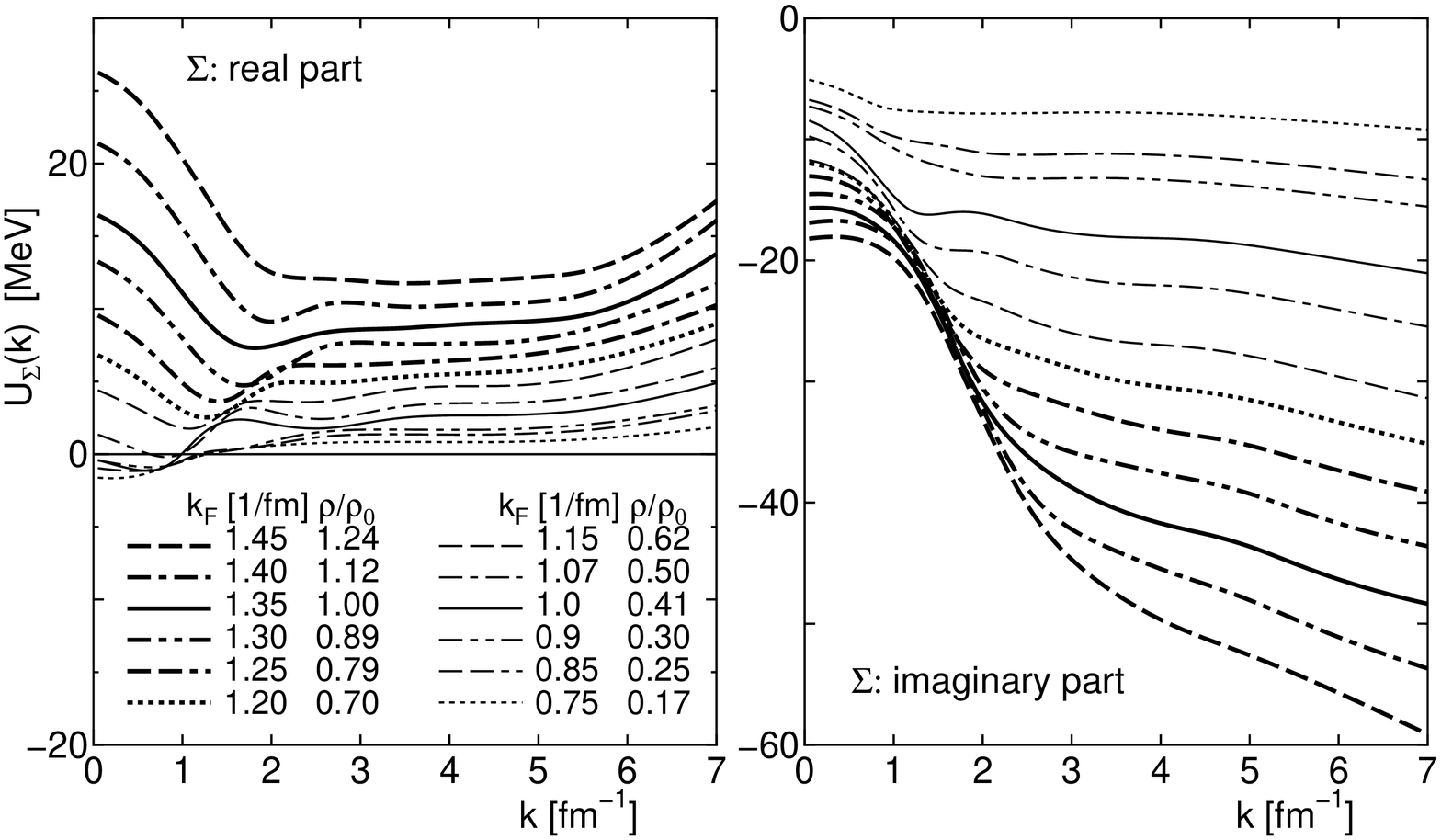}
\\
\medskip
 \epsfxsize=\textwidth
 \epsffile{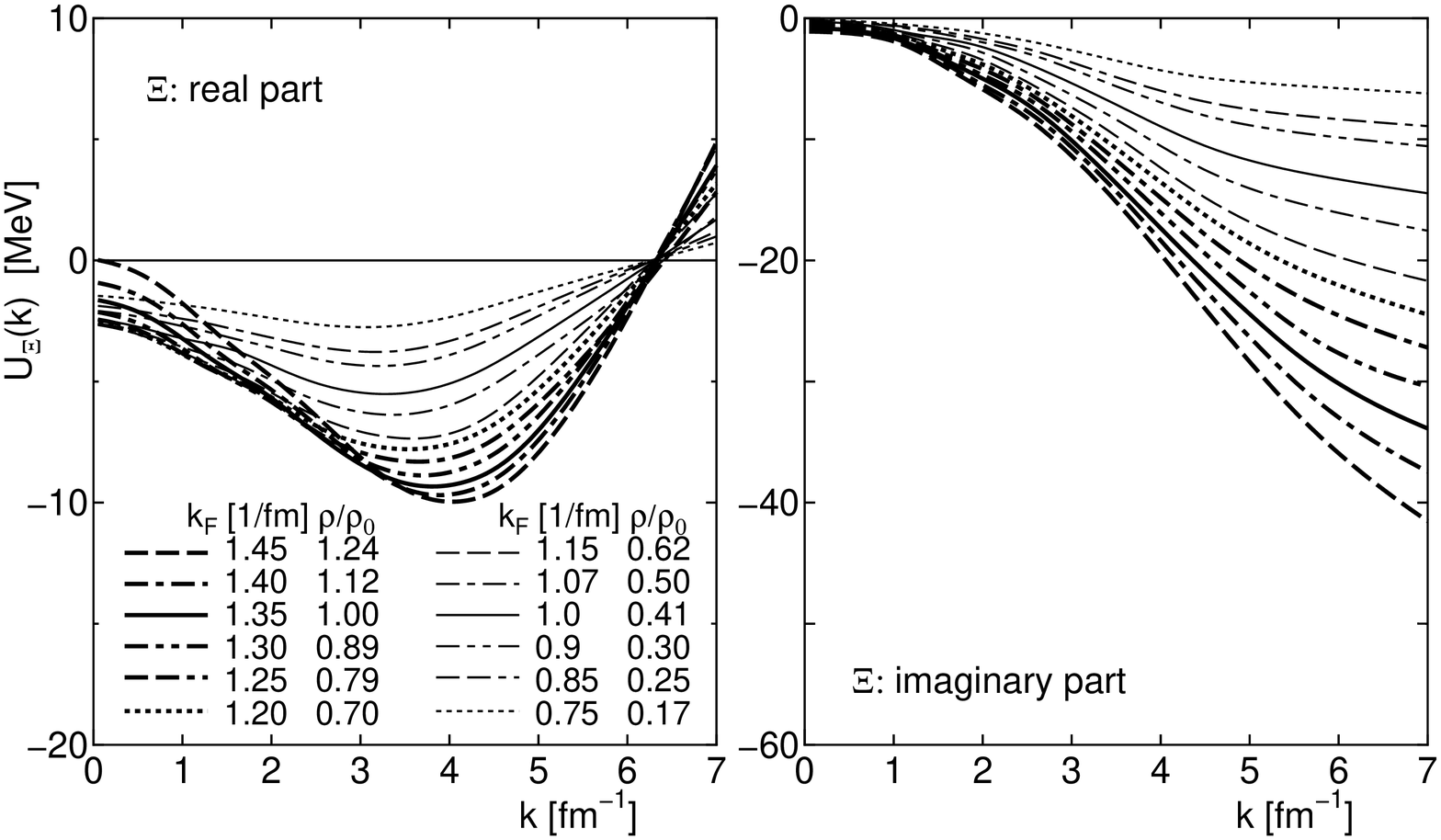}
}
\end{minipage}
\nolinebreak
\hspace*{5mm}
\begin{minipage}{0.26\textwidth}
{
 \caption{Real and imaginary parts of single-particle potentials for $\Sigma$
and $\Xi$ in symmetric nuclear matter at various
Fermi momenta, $0.75$ fm$^{-1}$ $\leq k_F \leq$ 1.45 fm$^{-1}$.}
}
\end{minipage}
\end{figure}
\end{widetext}

\noindent
tributions from higher order
diagrams and three-body forces are now semi-quantitatively understood in the
nucleon sector \cite{BA01}.

Real and imaginary parts of the calculated s.p. potentials for $N$, $\Lambda$,
$\Sigma$, and $\Xi$ in symmetric nuclear matter are shown in Figs.~1 and 2 as a function
of the momentum $k$. These are the results after the self-consistency for the
starting energy $\omega$ being reached. The Fermi momentum $k_F$ is chosen
approximately in a step of one tenth of the normal nucleon density $\rho_0$.
These densities are used as the discretized points of the density in the local
density approximation for considering finite nuclei. Below $k_F=0.75$ fm$^{-1}$,
the nuclear matter $NN$ $G$-matrix calculation becomes unstable due to the
appearance of a bound state in the $^3S_1$ channel. Because we expect little
relevance of this phenomenon to ground states of finite nuclei, the instability
is not inspected further. In the case that the effective interactions at low
density below $k_F=0.75$ fm$^{-1}$ is needed in the local density approximation
for finite nuclei, we use the $G$-matrices at $k_F=0.75$ fm$^{-1}$.

As for the nucleon, the result is very similar to those by other realistic
$NN$ potentials. The depth of the s.p. potential at the normal density is
considerably larger than the magnitude of the standard Woods-Saxon
potential, which is $50 \sim 60$ MeV. It has been known that the rearrangement
potential reduces the strength by about $10\sim 20$ MeV.

We are concerned mainly with the prediction of the quark-model
potential fss2 \cite{QMBB} for hyperon s.p. potentials. The strength of
the attractive $\Lambda$ s.p. potential in normal nuclear matter shown in Fig.~1
is almost 45 MeV, which is again larger than the empirically known
value of around 30 MeV. At the low density the potential is shallower. However,
as will be shown in the next section, the depth of the calculated $\Lambda$
s.p. potential in finite nuclei, taking into account the finite geometry and the
effects of the non-diagonal properties of the $G$,  seems to be dictated by
the potential depth at the normal nuclear density.
We can expect that the rearrangement effects give rise to a repulsive
contribution of the order of 10 MeV for the $\Lambda$ mainly through the energy
dependence of the $G$-matrix.

Nuclear matter calculations using the early version of the Kyoto-Niigata SU$_6$
quark-model potential, FSS, predicted a repulsive $\Sigma$ s.p.
potential \cite{KF00}. Results shown in Fig.~2 are obtained by the most recent
quark-model parameterization, fss2 \cite{QMBB}. The $\Sigma$ potential at $k=0$
is definitely repulsive of about 15 MeV at normal density. This repulsion chiefly
comes from the strong repulsive contribution of the $^3S_1$ state in the isospin
$T=\frac{3}{2}$ channel, which is naturally predicted by the quark-model
as the consequence of the Pauli principle on the quark level. The interaction
in the $^1S_0$ with $T=\frac{1}{2}$ channel is also repulsive. These
repulsive contributions overwhelm the attractive contributions from the $^1S_0$
with $T=\frac{3}{2}$ and the $^3S_1$ with $T=\frac{1}{2}$ channels.
The width $\Gamma$ of the $\Sigma$ hyperon in nuclear medium is related to the 
imaginary strength of the s.p. potential by $\Gamma(k) = -2 \Im U(k)$. $\Gamma(0)$
is seen in Fig.~2 to be more than 30 MeV at normal density.

The $\Xi$ s.p. potential in symmetric nuclear matter predicted by fss2 is weakly
attractive as is shown in Fig.~2. As the momentum increases, the magnitude of
the attraction is seen to become larger at the low momentum region
of $k < 6$ fm$^{-1}$. The momentum dependence may be characterized by
the effective mass. To obtain a rough estimation of it, we parameterize the
potential by $U_\Xi^{\mbox{\footnotesize real}}(k) \simeq ak^2+b$. In this
case the effective mass at $k=0$ is obtained by
\begin{equation}
 \frac{m_\Xi^*}{m_\Xi}=\left[1+\frac{2m_\Xi a}{\hbar^2}\right]^{-1}
\end{equation}
Calculated s.p. potentials give $m_\Xi^* /m_\Xi \sim 1.1$ at $k_F=1.35$ fm$^{-1}$
and $m_\Xi^* /m_\Xi \sim 1.05$ at $k_F=1.07$ fm$^{-1}$.
 
The $\Xi^- p$ elastic and inelastic cross-section measurements at low energy by
Ahn \textit {et al.} \cite{AHN06} indicate that the width of a $\Xi$ s.p. state
in nuclear medium is $\Gamma \sim 3$ MeV. Although it is uncertain at which
energy this number should be compared with the calculated imaginary strength,
the small imaginary strength of the $\Xi$ s.p. potential at the low momentum region
given in Fig.~2 is in accord with the empirical small width of the $\Xi$ in
nuclear medium.

It is encouraging to see that the results for $\Sigma$ and $\Xi$ hyperons
agree at least qualitatively with empirical indications so far obtained.

\section{Results in Finite Nuclei}

\begin{figure}[t!]
 \begin{center}
 \epsfxsize=0.45\textwidth
 \epsffile{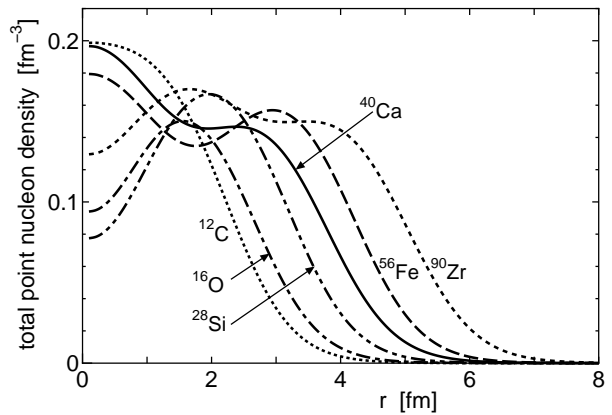}

 \caption{Point nucleon density distributions $\rho_t(r)$ obtained by DDHF
wave functions of the G-0 force \cite{CS}.}
\end{center}
\hfill
\end{figure}

\begin{figure}[t]
 \begin{center}
 \epsfxsize=0.4\textwidth
 \epsffile{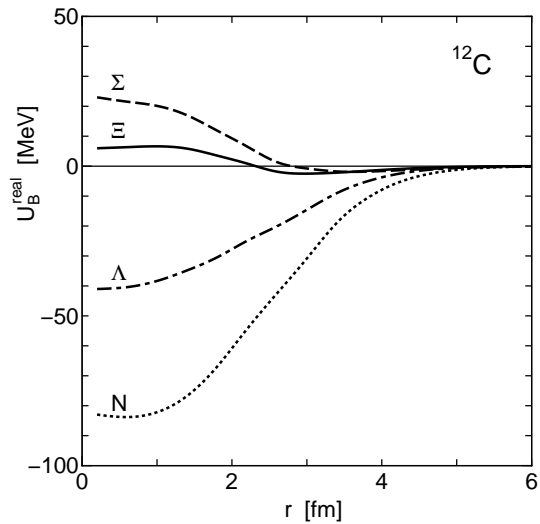}
 \caption{Real part of localized single-particle potentials for $N$, $\Lambda$,
 $\Sigma$, and $\Xi$ in $^{12}$C with the quark-model potential fss2 \cite{QMBB}
for the octet baryon-baryon interactions.}
\end{center}
 \end{figure}
 
\begin{figure}[t]
 \begin{center}
 \epsfxsize=0.4\textwidth
 \epsffile{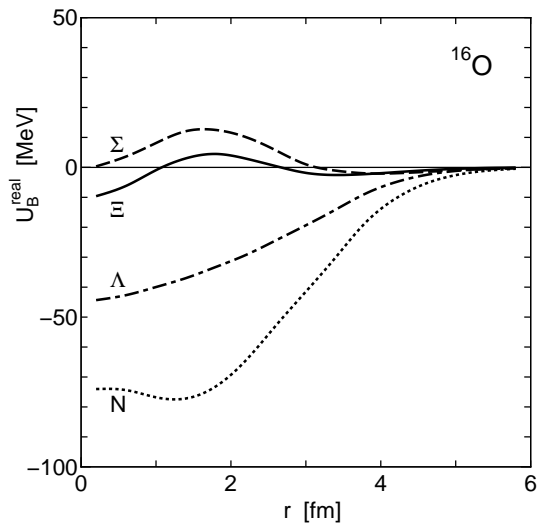}
 \caption{Same as in Fig.~4, but for $^{16}$O.}
\end{center}
 \end{figure}
 
 \begin{figure}[t]
 \begin{center}
 \epsfxsize=0.4\textwidth
 \epsffile{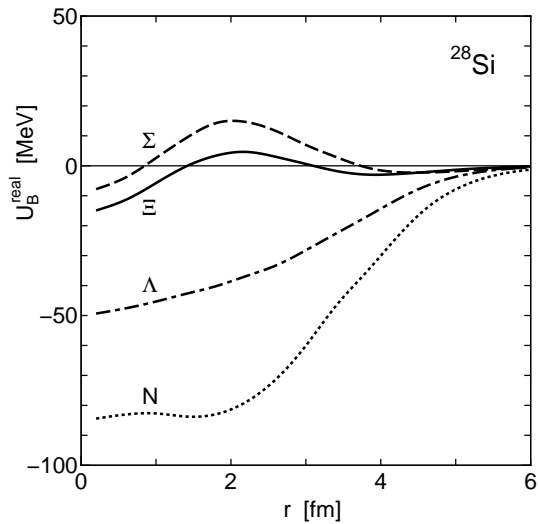}
 \caption{Same as in Fig.~4, but for $^{28}$Si.}
\end{center}
 \end{figure}
 
\begin{figure}[t]
 \begin{center}
 \epsfxsize=0.4\textwidth
 \epsffile{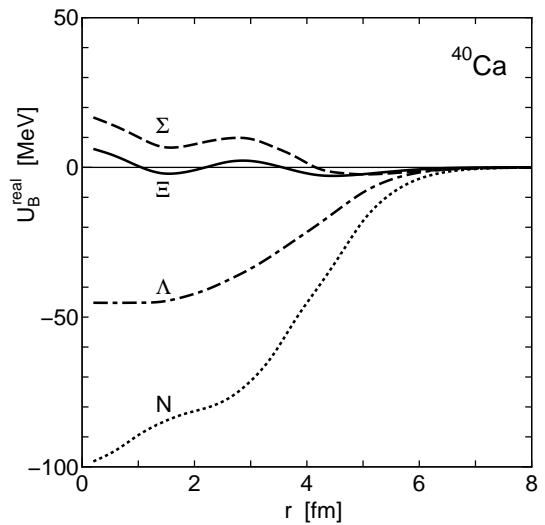}
 \caption{Same as in Fig.~4, but for $^{40}$Ca.}
 \end{center}
\end{figure}

\begin{figure}[t]
\begin{center} 
 \epsfxsize=0.4\textwidth
 \epsffile{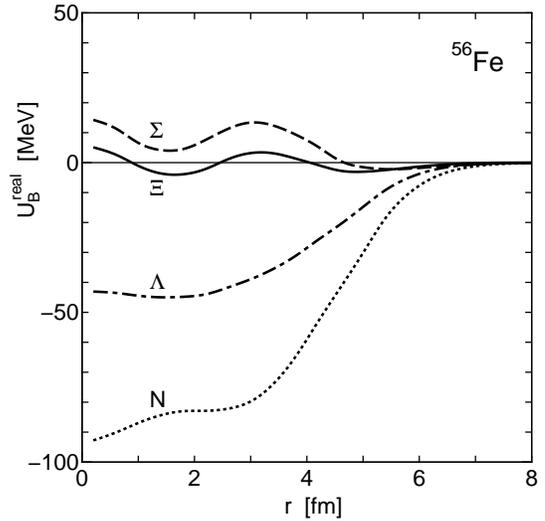}
 \caption{Same as in Fig.~4, but for $^{56}$Fe.}
\end{center}
 \end{figure}
 
 \begin{figure}[t]
 \begin{center}
 \epsfxsize=0.4\textwidth
 \epsffile{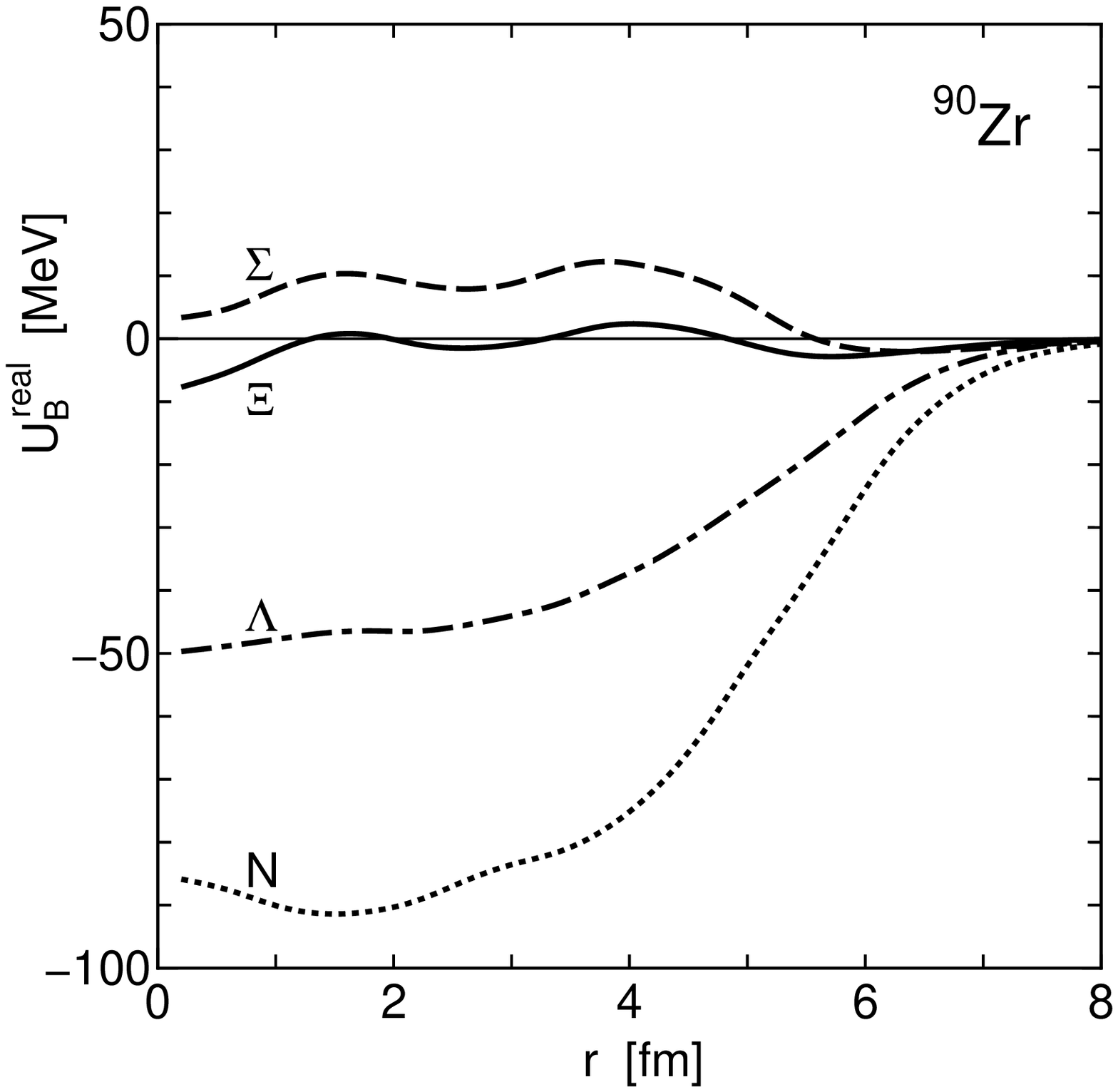}
 \caption{Same as in Fig.~4, but for $^{90}$Zr.}
\end{center}
 \end{figure}

We apply the calculational method presented in Sect. II to from light to
medium-heavy nuclei: $^{12}$C, $^{16}$O, $^{28}$Si, $^{40}$Ca, $^{56}$Fe,
and $^{90}$Zr. Nucleon density distributions are prepared by density-dependent
Hartree-Fock calculations using the Campi-Sprung G-0 force \cite{CS}. Profiles
of the point nucleon density distribution $\rho_t(r)$ which is a sum of the
neutron and proton densities are shown in Fig.~3.

Nuclear matter $G$-matrices are used in finite nuclei by the local density
approximation. At the position $R_1$ where the s.p. potential is evaluated
the local Fermi momentum is defined by the correspondence
$k_F(R_1)=\left[\frac{3\pi^2}{2}\rho_t(R_1)\right]^{1/3}$.
The $G$-matrices calculated in nuclear matter with this Fermi momentum
are used in Eq. (19). In actual calculations, $G$-matrix calculation is carried
out only for the Fermi momenta shown in Figs.~1 and 2. At each position $R_1$,
the Fermi momentum which is closest to $k_F(R_1)$ among these twelve values
is chosen. As explained in Sect. II, for small densities below
$k_F=0.75$ fm$^{-1}$, namely the total density $\rho_t=0.028$ fm$^{-3}$,
we always use $k_F=0.75$ fm$^{-1}$. In homogeneous matter the s.p. potential 
is determined by the matrix element with the zero momentum transfer, namely
diagonal ($\mbox{\boldmath{$k$}}'\:=\:\mbox{\boldmath{$k$}}$) components of
$G(\mbox{\boldmath{$k$}}',\mbox{\boldmath{$k$}};\mbox{\boldmath{$K$}},\omega)$.
In finite nuclei non-diagonal components of
the $G$ matrices also contribute to the s.p. potential.

The starting-energy dependence of the $G$-matrix plays an important role in
the LOBT. The prescription of the starting-energy as the sum of s.p. energies
of the two baryons considered means that certain higher-order diagrams are
included. Hence the self-consistency between the s.p. energy which is defined
by the $G$-matrix and the $G$-matrix which depends on the starting-energy is
required. Calculations in nuclear matter shown in Sect. II are the results
with this consistency achieved. In the case that the $G$-matrix in nuclear
matter is applied to a finite nucleus, however, there is no simple way to
treat the starting-energy dependence. We introduce an ad hoc prescription to
use an s.p. potential value at the median momentum $2^{-1/3}k_F$:
 $\omega = 2m_N+\frac{\hbar^2}{m_N}k^2
+\frac{\hbar^2}{4m_N}K^2+ U_N(2^{-1/3}k_F)+U_N(2^{-1/3}k_F)$
for the nucleon and $\omega = m_Y +m_N+\frac{\hbar^2( m_Y+ m_N)}{2m_Ym_N}k^2
+\frac{\hbar^2}{2(m_Y+m_N)}K^2+ U_Y(0) +U_N(2^{-1/3}k_F)$ for the hyperons.

The results with the $SU_6$ quark-model potential fss2 \cite{QMBB} in the
energy-renormalized form  are shown in Figs.~4$\sim$9. The charge state of
the baryon specified by $M_B$ in Eq. (19) is set to be $M_B=-I_B$. Comments
on the calculated s.p. potential of each baryon are given in the following.

\subsection{Nucleon s.p. potential}
The shape of the calculated nucleon potential is seen to follow the density
distribution, and the depth is $80\sim 90$ MeV which corresponds to the
s.p. potential in nuclear matter at the normal density. It is well known
that the straightforward application of the LOBT starting from realistic
$NN$ interactions overestimate the attractive nucleon-nucleus potential.
To compare the calculated potential with the empirical one, we need to
include the so-called rearrangement potential. The repulsive strength is
known to be $10\sim20$ MeV \cite{CS}. If this contribution is taken into
account, the resulting potential becomes closer to the phenomenological
potential of the Woods-Saxon form.

It is noted that it is still a remaining problem in nuclear physics to
understand nuclear bulk properties in a fully microscopic way on the
basis of the realistic interactions including higher-order correlations,
three-body forces, and other possible medium effects.

\subsection{$\Lambda$ hyperon s.p. potential}
The $\Lambda N$ $^1S_0$ state has a similar character to the $NN$ $^1S_0$ state
in the spin-flavor SU$_6$ symmetry, although there is small admixture of
a completely quark Pauli forbidden component. Similarly, the interaction
in the $\Lambda N$ $^3S_1$ channel resembles that in the $^3S_1$ $NN$ channel
with a smaller magnitude by a factor of $1/\sqrt{2}$, although there is an
important difference that the pion exchange is absent. Thus it is probable
that the $\Lambda$-nucleus potential is attractive, about a half of
the $N$-nucleus potential in magnitude. Looking at the density-dependence
of the $\Lambda$ s.p. potential in nuclear matter, we should expect a
similar rearrangement potential as in the nucleon case. The addition of the
hyperon to nuclear medium does not change directly the nucleon density and hence
the nucleonic Pauli effect. The rearrangement effect for the $\Lambda$ hyperon
originates, in the LOBT, mainly from the energy-dependence
of the $NN$ and $\Lambda N$ $G$-matrices. If we assume a repulsive
rearrangement potential of the order of 10 MeV, calculated results shown in
Figs.~4 $\sim$ 9 correspond well to the empirical $\Lambda$-nucleus potential
in the Woods-Saxon form with the depth of about 30 MeV.

\subsection{$\Sigma$ hyperon s.p. potential}
The experimental information has been limited for the $\Sigma$ s.p.
potential in nuclei. Because the $\Sigma$ state in a nucleus is expected
to have a large width due to the strong $\Sigma N\rightarrow \Lambda N$
conversion process, it is unlikely to observe clear peak structure in
the $\Sigma$-formation spectra. Nevertheless, results from the early
experiments of $(K, \pi)$ inclusive spectra \cite{BERT} measured at CERN
were interpreted as indicating that the $\Sigma$-nucleus potential is
moderately attractive. The discovery of the bound $^3$He+$\Sigma^0$ and
$^3$H+$\Sigma^+$ systems  \cite{NAGA} and the theoretical consideration
by Harada \textit{et al.} \cite{HARA98} showed that the attraction in
the $\Sigma N$ $T=1/2$ channel should be attractive enough. Another important
source of the $\Sigma$-nucleus interaction is the energy shift and the
width of $\Sigma^-$ atomic orbits extracted from the X-ray data.
Batty, Friedman, and Gal \cite{BFG94} analyzed the data to conclude that
the $\Sigma$-nucleus potential changes its sign toward higher density
region in a nucleus from the attractive potential at the surface region.
Dabrowski \cite{DA99} analyzed the BNL experiment of $(K,\pi)$ spectrum
on $^9$Be \cite{BNL} in a plane wave model and conjectured that the $\Sigma$
potential is repulsive of the order of $20$ MeV. Recent experimental
data with better accuracy of $(\pi^- ,K^+)$ inclusive spectra measured
at KEK \cite{KEK} was reported to suggest that the $\Sigma$-nucleus
potential is strongly repulsive, the strength being more than 100 MeV.
Several theoretical analyses carried out later \cite{HH05,K06} confirmed
the repulsive nature of the $\Sigma$ s.p. potential, but the height may
be a few 10 MeV. 

On the theoretical side, $\Sigma N$ interaction models admit large
uncertainties. Most parameter sets of the Nijmegen hyperon-nucleon
OBEP potential \cite{HCD,HCF,NSC,SR99} predict an attractive $\Sigma$-nucleus
potential in nuclear matter, except the model F \cite{YB85,YB90,SCHU}.
On the contrary, the strong repulsive character in the $T=3/2$ $^3S_1$
channel is inherent in the quark-model description \cite{QMBB,OSY}.
Thus the Kyoto-Niigata quark-model potential predicts an overall repulsive
$\Sigma$ s.p. potential in nuclear matter. Results given in Figs.~4$\sim$9
show the consequence of this property to finite nuclei. At higher
density region inside a nucleus the $\Sigma$-nucleus potential is
repulsive of 10$\sim$20 MeV. The overall repulsive nature of the
$\Sigma$-nucleus potential has been deduced by the analyses of
$(K^-,\pi^+)$ $\Sigma^-$ formation inclusive spectra \cite{KEK,HH05,K06}.
Beyond the surface region the potential becomes attractive. It is
interesting to see that the radial dependence indicated by the analyses
of $\Sigma$ atomic data \cite{BFG94} is actually derived by the
microscopic calculation using the quark-model bare interaction with
no phenomenological adjustment.

It is remarked that in the present evaluations we apply $G$-matrices in
symmetric nuclear matter to finite nuclei without separating neutron
and proton density distributions. However, in heavy nuclei, e.g. $^{90}$Zr
in our calculations, in which neutron and proton distributions are visibly
different, we should take care of the isospin-dependence when solving the
Bethe-Goldstone equation. In that case, the repulsive contribution to the
$\Sigma^-$ s.p. potential from the $T=3/2$ $^3S_1$ channel
becomes more predominant.

\subsection{$\Xi$ hyperon s.p. potential}
For the $\Xi$ s.p. potential, the experimental information has been more
scarce than the $\Sigma$. The $\Xi$ hyperon is formed in $(K^-,K^+)$
reaction on nuclei with small production rates. There has been no concrete
evidence of the $\Xi$ hypernuclear bound state. The existing experimental
data of the $\Xi$ formation spectrum \cite{II92,TF98,KH00} has suggested
that the $\Xi$ feels attractive potential in nuclei, the depth of which
is not so large, $10\sim 20$ MeV. Our results shown in Figs.~4$\sim$9
with the quark-model potential fss2 \cite{QMBB} show that
the $\Xi$-nucleus potential is weakly attractive at and beyond the nuclear
surface region, which is similar to the $\Sigma$-nucleus potential.
Toward the inside of the nucleus the $\Xi$ s.p. potential tends to be
repulsive and oscillates around zero with the magnitude of about 10 MeV.
It is not possible to simulate the potential shape by a single Woods-Saxon
form. No $\Xi$ hypernuclear bound state is expected from such a weak
potential. The situation does not change even if the actual potential
strength differs by a factor of 2 or so. In that case the level shift
of the $\Xi^-$ atomic orbit should be a valuable source of the information
about the $\Xi$-nucleus interaction. This subject is addressed in the
next section.

The evaluated $\Xi N$ $G$-matrices include full baryon-channel
couplings, namely the possible $\Xi N$-$\Lambda\Lambda$-$\Sigma\Sigma$
or $\Xi N$-$\Lambda\Sigma$-$\Sigma\Sigma$ couplings. It is helpful
to use equivalent interactions in low-momentum space to check the character
of the $\Xi N$ interaction and the effect of the baryon-channel coupling in
each spin and isospin state. Inspecting the matrix elements in ref. \cite{KEQ},
we see that the $\Xi N$ effective interaction from the fss2 in the $T=1$ channels
both in $^1S_0$ and $^3S_1$ are repulsive. The $T=0$ $^3S_1$ interaction is
very weak, and the $T=0$ $^1S_0$ interaction is attractive for which
the $\Xi N$-$\Lambda\Lambda$-$\Sigma\Sigma$ coupling is responsible.
It turns out that the net $s$-wave contribution is small and thus the
attractive $p$-wave contribution plays an important role to make the $\Xi$
s.p. potential to be attractive at the surface region.

\section{Energy shift and width of atomic orbit}

\begin{table}
\caption{The strength and geometry parameters of the Woods-Saxon form
$f_i(r)=U_i^0 /[1+\exp((r-r_{0,i})/a_i)]$ fitted to the real part as well
as the imaginary part of $\Sigma$ and $\Xi$ s.p. potentials $U_B(r)$ in
$^{28}$Si and $^{56}$Fe.}
\begin{tabular}{ccr@{}c@{}lclr@{}c@{}lcl} \hline\hline
  & & \multicolumn{5}{c}{real part} & \multicolumn{5}{c}{imaginary part} \\ \hline
  & & \multicolumn{5}{c}{$\sum_{i=1,3} f_i(r)$}
 & \multicolumn{5}{c}{$\sum_{i=1,3} f_i(r)$} \\
 $^{28}$Si & $i$ & \multicolumn{3}{c}{$U_i^0$} & $r_{0,i}$ & $a_i$\hspace{1em}
 & \multicolumn{3}{c}{$U_i^0$} & $r_{0,i}$ & \hspace{1em}$a_i$ \\
   &  & \multicolumn{3}{c}{[MeV]} & [fm] & [fm]\hspace{0.5em}
   & \multicolumn{3}{c}{[MeV]} & [fm] & \hspace{0.5em}[fm] \\ \hline
      & 1 & $-25$&.&94 & 4.179 & 0.7164  & $-65$&.&63 & 3.819~ & 0.7539 \\
 $\Sigma$  & 2 & +57&.&43 & 3.049 & 0.7860 & +41&.&93 & 3.997~ & 0.8185 \\
      & 3 & $-41$&.&13 & 1.220 & 0.4348 & $-6$&.&078 & 0.3944 & 1.576 \\ \hline
      & 1 & $-310$&.&4 & 2.171 & $1.066\;\,$ & $-6$&.&760 & 4.980~ & 0.6915 \\
 $\Xi$  & 2 & +543&.&3 & 2.959 & 0.9484 & +5&.&639 & 5.209~ & 0.6668 \\
    & 3 & $-270$&.&3 & 3.421 & 0.8630 & $-1$&.&838 & 1.118~ & 0.2948 \\ \hline
  & & \multicolumn{5}{c}{real part} & \multicolumn{5}{c}{imaginary part} \\ \hline
  & & \multicolumn{5}{c}{$\sum_{i=1,2} f_i(r) +\frac{df_3(r)}{dr}$}
 & \multicolumn{5}{c}{$\sum_{i=1,3} f_i(r)$} \\
 $^{56}$Fe & $i$ & \multicolumn{3}{c}{$U_i^0$}
 & $r_{0,i}$ & $a_i$\hspace{1em} &  \multicolumn{3}{c}{$U_i^0$} & $r_{0,i}$
 & \hspace{1em}$a_i$ \\
 &  & \multicolumn{3}{c}{[MeV]} & [fm] & [fm]\hspace{0.5em}
 & \multicolumn{3}{c}{[MeV]} & [fm] & \hspace{0.5em}[fm] \\ \hline
  & 1 & $-3$&.&746 & 6.035 & 0.5655& $-65$&.&78 & ~4.955 & 0.8902 \\
 $\Sigma$  & 2 & 22&.&19 & 4.031 & 0.3990 & +37&.&65 & ~5.486 & 0.8867 \\
  & 3 & 32&.&22 & 1.553 & 0.5527 & +16&.&77 & $-0.4167$ & 1.163 \\ \hline
  & 1 & $-2$&.&232 & 6.367 & 0.2389 & $-15$&.&92 & ~5.845 & 0.6820 \\
 $\Xi$ & 2 & +8&.&295 & 3.796 & 0.2597 & +14&.&64 & ~5.947 & 0.6728 \\
  & 3 & +20&.&67 & 1.674 & 0.4887 & +1&.&290 & ~~0.7581 & 0.1709 \\ \hline \hline                         
\end{tabular}

\end{table} 

The level shift and the width of $\Sigma^-$ atomic orbits are a valuable source
of the information on the $\Sigma$-nucleus strong interaction. The analyses
by Batty, Friedman, and Gal \cite{BFG94} indicated that the $\Sigma$-nucleus
potential is attractive at the surface region, but at higher density region
in a nucleus the potential turns to be repulsive. The radial dependence of
the calculated $\Sigma^-$ s.p. potential shown in the previous section agrees
with this. Therefore it is instructive to explicitly evaluate the energy
shift and the width of $\Sigma^-$ atomic orbits with the calculated potential.
As will be shown below, the result is consistent with the experimental data.
This indicates that the microscopic calculation with the quark-model fss2 is
reliable in the $\Sigma N$ channel. Thus, it is interesting to extend the
level shift calculation to $\Xi^-$ atomic orbits. The experimental data
should be available in near future, because the first measurement of $\Xi^-$
atomic X rays from Fe target is proposed \cite{TANI} to be performed at J-PARC.
The theoretical prediction provides a guiding information for this experiment.

We consider $^{28}$Si and $^{56}$Fe for explicit evaluations of the level
shift of the atomic orbit.  We first fit the shape of the calculated s.p.
potential using the Woods-Saxon form. For  $^{28}$Si a sum of three Woods-Saxon
shapes is used and for $^{56}$Fe a sum of two Woods-Saxon shapes and one
derivative of the Woods-Saxon shape is assigned. Parameters are given in Table I.
It is noted that the imaginary parts are also given to illustrate the order of
the magnitude of the absorptive strength, intending to demonstrate that the $\Xi$
imaginary potential is about one order of magnitude smaller than the $\Sigma$ one.
However, actual numbers should not be taken very seriously because nuclear
matter calculations tend to overestimate the imaginary strength as the
calculations \cite{CEG} of nucleon optical model potential indicates.
In addition, the prescription to use the $G$-matrices at $k_F=0.75$ fm$^{-1}$
for all the densities below $k_F=0.75$ fm$^{-1}$ probably leads to the
overestimation of the imaginary strength at the surface region. It is also
remarked that localized imaginary potential through the zero-momentum Wigner
transformation may become positive at some points.

\subsection{$\Sigma^-$}
Results of the level shift $\Delta E=E-E_C$ and the width $\Gamma =-2\Im E$
for the $\Sigma^-$ $f$- and $g$-atomic levels on $^{28}$Si and the
$\Sigma^-$ $g$- and $h$-atomic levels on $^{56}$Fe are given in Table II,
where $E_C$ stands for the Coulomb bound state energy without
the $\Sigma^-$-nucleus strong interaction. When the real part of the
$\Sigma$ s.p. potential is taken into account, the energy of the
$n=4, \ell=3$ orbit on $^{28}$Si is shifted downward by 222 eV. To
investigate the contribution of the absorptive effect, we do not use the
calculated potential given in Table I. The imaginary potential is rather
strong as explained above. The magnitude of the level shift and the width
depends non-linearly on the  strength of the imaginary potential. Hence we
use an phenomenological imaginary potential to discuss the level shift of the
atomic orbits of the $\Sigma^-$. We add an imaginary potential in the single
Woods-Saxon form used in ref. \cite{BFG94}, namely $r_0=1.1\times 28^{1/3}$ and
$a=0.67$ with the depth of $-9$ MeV. In that case, we obtain $\Delta E=208$ eV
and $\Gamma =249$ eV, which well correspond to the experimental values of
$\Delta E_{exp}=159\pm 36$ eV and $\Gamma_{exp}=220\pm 110$ eV. This result
indicates that the real part of the $\Sigma$ s.p. potential calculated
microscopically in the LOBT starting from the two-body quark-model potential
fss2 is reasonable.

\begin{table}
\caption{The energy shift $\Delta E=E-E_C$ and the width $\Gamma=-2 \Im E$ of
the $\Sigma^-$ atomic orbits in $^{28}$Si and  $^{56}$Fe, using the parameterized $\Sigma$
s.p. potential given in Table I for the real part. The imaginary potential is
given in a Woods-Saxon form with the strength $W_0=-9$ MeV and the geometry
parameters $r_0=1.1 A^{1/3}$ fm and $a=0.67$ fm. Entry numbers are in eV.}
\begin{tabular}{ccc} \hline\hline
 \multicolumn{3}{c}{$\Sigma^-$-$^{28}$Si} \\ \hline
 \hspace*{10em} & \hspace{2em}$\Delta E_{4f}$\hspace{2em}
 & \hspace{2em}$\Gamma_{4f}$\hspace{2em} \\
  real part only       & $-222$ & --- \\
  real + imaginary & $-208$ & 249 \\
  exp. \cite{B78}   & $-159\pm 36$ & $220\pm 110$ \\ \hline
                          & $\Delta E_{5g}$ & $\Gamma_{5g}$ \\
  real part only       & $-0.8$ &--- \\
  real + imaginary  & $-0.8$ & 0.7 \\
  exp. \cite{B78}   &    ---    & $0.41\pm 0.1$     \\ \hline
 \multicolumn{3}{c}{$\Sigma^-$-$^{56}$Fe} \\ \hline
                         & $\Delta E_{5g}$ & $\Gamma_{5g}$ \\
  real part only       & $-943$ & --- \\
  real + imaginary  & $-943$ & 1205 \\ \hline
                          & $\Delta E_{6h}$ & $\Gamma_{6h}$ \\
  real part only       & $-11$ & --- \\
  real + imaginary  & $-11$ & 8.3 \\ \hline \hline
\end{tabular}
\end{table}

\subsection{$\Xi^-$}
Observing that our calculated $\Sigma$-nucleus potential gives a reliable
result for the shift of the $\Sigma^-$ atomic level, it is interesting to
proceed to $\Xi^-$ atoms without any adjustment. In ref. \cite{BFG99} Batty,
Friedman, and Gal estimated the level shift and the width of $\Xi^-$ atoms for
the $\Xi^-$-nucleus potential having an attraction of the depth of 15-20 MeV
with an imaginary strength of 1-3 MeV in the shape roughly following
a nuclear density distribution. Although the $\Xi$-nucleus potential is
weakly repulsive inside the nucleus in our calculation, the attractive
strength at the surface region is found to be comparable to that of the
Woods-Saxon potential with the depth of $10\sim 20$ MeV and the geometry
parameters of $r_0=1.1A^{1/3}$ fm and a=0.67 fm. Results for
the $f$- and $g$-orbits in $^{28}$Si and the $g$- and $h$-orbits in
$^{56}$Fe are given in Table III, together with those of the Woods-Saxon
potential as a reference. Because the atomic level shift is insensitive to
the short range part of the strong interaction potential, our potential
parameterized as shown in Table I, predicts a similar magnitude of the
level shift of the reference Woods-Saxon potential. Because the $\Xi$
imaginary potential obtained by the $G$-matrix is small, we directly
use it for the estimation of the width $\Gamma$, though it is likely
to overestimate the absorptive effect as in the case of $\Sigma^-$.
As is seen in Table III, the width is of the order of a few hundred eV
for the g-orbit in $^{56}$Fe and the energy shift is hardly affected.
It is needed to see what order of the magnitude is detected for the
$\Xi^-$ $n=5, \ell=4$ level in $^{56}$Fe in the future experiment
prepared at J-PARC \cite{TANI}.

\begin{table}
\caption{The energy shift $\Delta E=E-E_C$ and the width $\Gamma=-2 \Im E$
of the $\Xi^-$ atomic orbits in $^{28}$Si and $^{56}$Fe, using the parameterized $\Xi$
s.p. potential given in Table I both for the real and imaginary parts.
As a reference, results obtained by the complex Woods-Saxon potential
with the strength of $U_0=-14-3 i$ MeV, $r_0=1.1 A^{1/3}$ fm, and a=0.67 fm
are included. Entry numbers are in eV.}
\begin{tabular}{ccc} \hline\hline
 \multicolumn{3}{c}{$\Xi^-$-$^{28}$Si} \\ \hline
 \hspace*{10em} & \hspace{2em}$\Delta E_{4f}$\hspace{2em}
  & \hspace{2em}$\Gamma_{4f}$\hspace{2em} \\
  real part only       & $-346$ & --- \\
  real + imaginary  & $-345$ & $\:\;16$ \\
  reference pot.  & $-383$ & 216 \\ \hline
                          & $\Delta E_{5g}$ & $\Gamma_{5g}$ \\
  real part only       & $-6.9$ & --- \\
  real + imaginary  & $-7.0$ & 0.0 \\
  reference pot.  & $-1.4$ & 0.5 \\ \hline
 \multicolumn{3}{c}{$\Xi^-$-$^{56}$Fe} \\ \hline
                         & $\Delta E_{5g}$ & $\Gamma_{5g}$ \\
  real part only       & $-1287$ & --- \\
  real + imaginary  & $-1281$ & $\quad 88$ \\
  reference pot. & $-1675$ & 1092 \\  \hline
                          & $\Delta E_{6h}$ & $\Gamma_{6h}$ \\
  real part only       & $-12$ & --- \\
  real + imaginary  & $-12$ & 1.0 \\
  reference pot.   & $-17$ &  8.0 \\ \hline \hline
\end{tabular}
\end{table} 

\section{Summary}
In order to examine the prediction of single-particle properties of
all the octet baryons in nuclear medium, especially $\Xi$ hyperon,
by the recently developed quark-model baryon-baryon interactions,
we have evaluated localized s.p. potentials in finite nuclei by
folding $G$-matrices in nuclear matter with respect to nucleon
s.p. wave functions in the scheme of the local density approximation.
Introducing a spin-average approximation and a zero-momentum Wigner
transformation, the non-local baryon s.p. potential calculated in
momentum space is reduced to a local potential in coordinate space.
The final expression is feasible for numerical calculations, in the
case that the nucleon density distribution is discretized. Adopting
about one tenth of the normal nuclear density as the interval of the
discretization, we have carried out calculations
in $^{12}$C, $^{16}$O, $^{28}$Si, $^{40}$Ca, $^{56}$Fe, and $^{90}$Zr
for each octet baryon; N, $\Lambda$, $\Sigma$, and $\Xi$. This is the
first comprehensive evaluations of the s.p. potentials of all the octet
baryons in finite nuclei, starting from the baryon-baryon bare
interactions. These microscopic calculations of octet baryon s.p.
potentials in finite nuclei are meaningful to elucidate the character
of the theoretical model of the octet baryon-baryon interactions by
comparing them with empirical s.p. potentials, which are not yet
available for the $\Sigma$ and $\Xi$ hyperons. 

We use the most recent quark-model potential, fss2 \cite{QMBB} as bare
baryon-baryon interactions. The energy-dependence in the original form
of the quark-model potential is eliminated by the renormalization procedure.
The $NN$ sector of this potential describes scattering data as accurately as
other modern realistic interaction. Calculated nucleon s.p. potentials in
nuclear medium are found to be similar to those obtained in the LOBT
framework with other potentials. The $\Lambda N$ interaction is under
control to a certain extent by the experimental data of $\Lambda$
hypernuclei. The fss2 gives similar $\Lambda$ s.p. potentials in nuclear
medium to those of the Nijmegen OBEP potential. It was shown \cite{KEQ}
in fact that the fss2 and the Nijmegen NSC97f actually have very similar
matrix elements of the equivalent interaction in low-momentum space.

The extension to the $\Sigma N$ channel and further to the $\Xi N$
sector of the strangeness $S=-2$ has many ambiguities because of
scarcity of experimental information. It is necessary to rely on the
theoretical framework as reliable as possible to construct these baryon-baryon
interactions. At present, the SU$_6$ quark-model fss2 is more predictive
than the OBEP model, in the sense that the potential parameters are
uniquely given in contrast to various sets of parameters presented by
the Nijmegen group. Thus we focus our attention in this paper on the 
fss2 as the input $\Sigma N$ and $\Xi N$ two-body interactions.
The comparison with the results by other potential models including
the parameterization based on the effective chiral
field approach \cite{EFT1,EFT2} is an interesting future subject.

Properties of the calculated $\Sigma$ s.p. potential in finite nuclei
are found to agree well with the experimental evidence so far obtained.
The potential is repulsive of the order of 10-20 MeV, which is necessary
to accounts for the $(\pi^-,K^+)$ $\Sigma^-$ formation spectra on
nuclei \cite{HH05,K06}, while it should be attractive at the surface
region, as the atomic level shifts indicate \cite{BFG94}. The attraction
obtained by the present microscopic calculation can reproduce the
empirical energy shifts of the $\Sigma^-$ atomic levels. To determine
more precisely the shape and the strength of the $\Sigma$-nucleus
potential including its isospin-dependence and study the relation to
the underlying bare interaction, we need further experimental data.

On the basis of these observations, it is interesting to consider
the $\Xi N$ sector. The imaginary part of the $\Xi$ s.p. potential
is small in the quark model description, which is consistent with
the empirical estimation based on the $\Xi^- p$ scattering
cross-section at low energy \cite{AHN06}. Therefore, if the real
part is sufficiently attractive to sustain hypernuclear bound states,
we can expect clear spike structure in the $(K^-,K^+)$ inclusive
spectra on nuclei. At the surface region the potential has similar
attraction to the $\Sigma$ potential. However, the potential does not
have a familiar shape simulated well by a Woods-Saxon form.
Inside the nucleus the potential fluctuates around zero, reflecting
the fluctuation of the nucleon density distribution. The quark model
potential fss2 implies that the net $s$-wave contribution to
the $\Xi$ s.p. potential is small and the net weakly attractive
contribution from the $p$-waves is relatively important. Such a
potential does not support $\Xi$ nuclear bound states. In that case,
measurements of the energy shift and width of the $\Xi$ atomic level
become an invaluable source of the information about the $\Xi$-nucleus
strong interaction. Our $\Xi$ s.p. potential calculated by the fss2
suggests that the negative energy shift is the order of a few hundred
eV for the $4f$-orbit in $^{28}$Si and the order of 1 keV for
$5g$-orbit in $^{56}$Fe. The existence or non-existence of $\Xi$
hypernuclear bound states and the energy shifts of the $\Xi^-$ atomic
states will be clarified in near future by the experiments prepared
in the J-PARC project \cite{JPARC},
which should advance our understanding of the interactions in the $S=-2$ sector.
Our calculational framework provide a useful method to link the baryon s.p.
properties in nuclei with the two-body interactions between octet baryons.

\bigskip
\acknowledgments

This study was supported by Grant-in-Aids for Scientific
Research (C) from the Japan Society for the Promotion of
Science (Grants nos. 17540263 and 18540261).


\end{document}